\documentclass[12pt]{article}

\newcommand{\be}{\begin{equation}}
\newcommand{\ee}{\end{equation}}
\newcommand{\bi}[1]{\vspace{-3mm} \bibitem{#1}}
\usepackage{epsfig,amsmath,amssymb,graphics,graphicx} 

\usepackage{graphics,graphicx}
\usepackage{amssymb,amsmath}
\usepackage{hyperref}
\usepackage{amsfonts}
\usepackage{epsfig}

\begin{document}

%%%%%%%%%%%%%%%%%%%%%%%%%%%%%%%%%%%%%%%%%%%%%%%%%%%%%%%%%%%%%%%%%%%%%%%
\begin{center}

{\it International Journal of Solids and Structures.
Vol.51. No.15-16. (2014) 2900-2907.} 

\vskip 7mm
{\bf \large Lattice with Long-Range Interaction of Power-Law Type    
\vskip 3mm
for Fractional Non-Local Elasticity} \\

\vskip 3mm
{\bf \large Vasily E. Tarasov} \\
\vskip 3mm

{\it Skobeltsyn Institute of Nuclear Physics,\\ 
Lomonosov Moscow State University, Moscow 119991, Russia} \\
{E-mail: tarasov@theory.sinp.msu.ru} \\

\begin{abstract}
Lattice models with long-range interactions of power-law type
are suggested as a new type of microscopic model for 
fractional non-local elasticity. 
Using the transform operation, we map the lattice equations
into continuum equation with Riesz derivatives of  non-integer orders.
The continuum equations that are obtained from the lattice model 
describe fractional generalization of 
non-local elasticity models.
Particular solutions and correspondent asymptotic 
of the fractional differential equations 
for displacement fields are suggested for the static case.
\end{abstract}

\end{center}

%%% 45.10.Hj Perturbation and fractional calculus methods
%%% 62.20.Dc Elasticity, elastic constants.
%%% 61.50.Ah Theory of crystal structure, crystal symmetry; calculations and modeling
%%% 81.40.Jj Elasticity and anelasticity, stress-strain relations

%%%%%%%%%%%%%%%%%%%%%%%%%%%%%%%%%%%%%%%%%%%%%%%%%%%%%%%%%%%%%%%

\section{Introduction}

Lattice with long-range interaction is a subject 
of investigations in different areas of mechanics and physics
(see for example \cite{Kro,EK1977,MOS2002,LA,TarasovSpringer} and
\cite{Dyson3,CMP,NakTak4,CDR}). 
The long-range interactions have been studied 
in discrete systems as well as in their continuous analogues.  
As it was shown in \cite{JPA2006,JMP2006} 
(see also \cite{CNSNS2006,TarasovSpringer}), 
the continuum equations with derivatives of non-integer orders 
can be directly connected to
lattice models with long-range interactions of power law type. 

The theory of derivatives and integrals of non-integer orders \cite{SKM,KST} 
allow us to investigate the behavior of materials and media 
that are characterized by non-locality of power-law type. 
Fractional calculus has a wide application in mechanics 
and physics (for example see 
\cite{CM,Hilfer,SATM,Mainardi,LA,TarasovSpringer,KLM,IJMPB2013}).
The fractional calculus allows us to formulate a
fractional generalization of non-local elasticity models 
in two forms:
the fractional gradient elasticity models (weak power-law non-locality) and 
the fractional integral non-local models (strong power-law non-locality).
Fractional models of non-local elasticity 
and some microscopic models are considered in different articles  
(see for example \cite{Laz,CCS-2,CPZ-2,CPZ-3,CPZ-4,CPZ-5}, 
\cite{PZ-1,PZ-2,PZ-3,PZ-4} and \cite{Arxiv2013,CEJP2013,MOM2014})
Elastic waves in nonlocal continua 
modelled by a fractional calculus approach are considered in 
\cite{Waves-1.Waves-2,Waves-3,Waves-4,Waves-5}.
In \cite{Arxiv2013,CEJP2013} a general approach to describe
lattice model with power-law spatial dispersion 
for fractional elasticity has been proposed.
This approach can be used for different type of interaction of lattice particles. Therefore
explicit forms of the long-range interactions are not considered in \cite{Arxiv2013,CEJP2013}.
In \cite{MOM2014} a model of lattice with 
long-range interaction of Gr\"unwald-Letnikov-Riesz type
has been suggested to describe
fractional gradient and integral elasticity of continuum.
In this paper we focus on the lattice models with long-range interaction of power-law type
as new type of microscopic models for
fractional generalization of elasticity theory.
We suggest lattice models with power-law long-range interaction
as microscopic model of fractional non-local continuum. 
The equations for displacement field 
are directly derived from the suggested
lattice models by the methods of \cite{JPA2006,JMP2006}. 
The suggested generalization of the elasticity equations
contains the fractional Laplacian in the Riesz's form \cite{KST}.
We demonstrate a connection between the dynamics of 
lattice system of particles with long-range interactions 
and the fractional continuum equations by using 
the transform operation suggested in \cite{JPA2006,JMP2006}. 
We show how the continuous limit for the lattice with long-range interactions of power-law type
gives the continuum equation of the fractional elasticity.
We get particular solutions of the fractional differential elasticity equations 
for some special cases.

%%%%%%%%%%%%%%%%%%%%%%%%%%%%%%%%%%%%%%%%%%%%%%%%%%%%%%%%%%%%%%%

\section{Equations of lattice model}

As a microscopic model, we use unbounded homogeneous lattices, 
such that all particles are displaced from its equilibrium positions 
in one direction, and the displacement of particle 
is described by a scalar field. 
We consider one-dimensional lattice system of interacting particles. The equations of motion for particles are
\be \label{Main_Eq2}
M \frac{d^2 u_n(t)}{d t^2} = 
g_2 \, \sum_{\substack{m=-\infty \\ m \ne n}}^{+\infty} \; K_2(n,m) \; \Bigl( u_n-u_m \Bigr) + 
g_{\alpha} \, \sum_{\substack{m=-\infty \\ m \ne n}}^{+\infty} \; 
K_{\alpha}(n,m) \; \Bigl( u_n-u_m \Bigr) + F (n) ,
\ee
where $u_n(t)$ are displacements from the equilibrium, 
$g_2$ and $g_{\alpha}$ are the coupling constants of particle interactions,  
and the terms $F(n)$ characterize an interaction of the particles   
with the external on-site force. 
For simplicity, we assume that all particles have the same mass $M$. 
The function $K_2(n,m)$ describes the nearest-neighbor interaction 
with coupling constant $g_2=K$, which is the spring stiffness. 
The function $K_{\alpha}(n,m)$ describes the long-range interaction 
with a coupling constant $g_{\alpha}$. 
For a simple case each particle can be considered an inversion center and 
\[ K_{\alpha}(n,m) = K_{\alpha}(|n-m|) . \]

Equations of motion (\ref{Main_Eq2}) have the invariance 
with respect to its displacement of lattice as a whole
in case of absence of external forces. 
It should be noted that the non-invariant terms lead to the divergences 
in the continuous limit \cite{TarasovSpringer}.

%%%%%%%%%%%%%%%%%%%%%%%%%%%%%%%%%%%%%%%%%%%%%%%%%%%%

Using the approach suggested in \cite{JMP2006,JPA2006,TarasovSpringer},
we can consider a set operations  that transforms
the lattice equations for $u_n(t)$ 
into continuum equation for displacement field $u(x,t)$. 
We assume that $u_n(t)$ are Fourier coefficients
of the field $\hat{u}(k,t)$ on $[-k_0 /2, k_0 /2]$ 
that is described by the equations
\be \label{un} 
u_n(t) = \frac{1}{k_0} \int_{-k_0/2}^{+k_0/2} dk \ \hat{u}(k,t) \; e^{i k x_n}= 
{\cal F}^{-1}_{\Delta} \{ \hat{u}(k,t) \} , 
\ee
\be \label{ukt}
\hat{u}(k,t) = \sum_{n=-\infty}^{+\infty} \; u_n(t) \; e^{-i k x_n} =
{\cal F}_{\Delta} \{u_n(t)\} ,
\ee
where $x_n = n d $ and $d =2\pi/k_0$ is 
distance between equilibrium positions of the lattice particles. 
Equations (\ref{ukt}) and (\ref{un}) are the basis for 
the Fourier series transform ${\cal F}_{\Delta}$ and the inverse 
Fourier series transform ${\cal F}^{-1}_{\Delta}$.

The Fourier transform can be derived from (\ref{ukt}) and (\ref{un}) 
in the limit as $d  \to 0$ ($k_0 \to \infty$). 
In this limit the sum is transformed into an integral, and 
equations (\ref{un}) and (\ref{ukt}) become
\be \label{ukt2} 
\tilde{u}(k,t)=\int^{+\infty}_{-\infty} dx \ e^{-ikx} u(x,t) = 
{\cal F} \{ u(x,t) \}, 
\ee
\be \label{uxt}
u(x,t)=\frac{1}{2\pi} \int^{+\infty}_{-\infty} dk \ e^{ikx} \tilde{u}(k,t) =
 {\cal F}^{-1} \{ \tilde{u}(k,t) \} . 
\ee
Here we use the lattice function
\[ u_n(t) = \frac{2 \pi}{k_0} u(x_n,t) \] 
with continuous function $u(x,t)$, where $x_n=nd = (2 \pi n) / k_0 \to x$. 
We assume that $\tilde{u}(k,t)= {\cal L} \hat{u}(k,t)$, 
where ${\cal L}$ denotes the passage 
to the limit $d  \to 0$ ($k_0 \to \infty$),
i.e. the function $\tilde{u}(k,t)$ can be derived from $\hat{u}(k,t)$
in the limit $d  \to 0$.
Note that $\tilde{u}(k,t)$ is a Fourier transform of the field $u(x,t)$.
The function $\hat{u}(k,t)$ is a Fourier series transform of $u_n(t)$,
where we can use $u_n(t)=(2\pi/k_0) u(nd ,t)$.

We can state that a lattice model transforms into continuum model 
by the combination ${\cal F}^{-1} {\cal L} \ {\cal F}_{\Delta}$ 
of the following operation \cite{JMP2006,JPA2006}: \\
The Fourier series transform:
\be \label{O1}
{\cal F}_{\Delta}: \quad u_n(t) \to {\cal F}_{\Delta}\{ u_n(t)\}=
\hat{u}(k,t) .
\ee
The passage to the limit $d  \to 0$:
\be \label{O2}
{\cal L}=\lim_{d \to 0}: \quad \hat{u}(k,t) \to {\cal L} \{\hat{u}(k,t)\}=
\tilde{u}(k,t) . \ee
The inverse Fourier transform: 
\be \label{O3}
{\cal F}^{-1}: \quad \tilde{u}(k,t) \to 
{\cal F}^{-1} \{ \tilde{u}(k,t)\}=u(x,t) .
\ee
These operations allow us to get continuum equations
from the lattice equations \cite{JMP2006,JPA2006,TarasovSpringer}.

%%%%%%%%%%%%%%%%%%%%%%%%%%%%%%%%%%%%%%%%%%%%%%%%%%%%%%%%%%%%%%%%%%%%%%
\section{Lattice with nearest-neighbor interaction}

Let us consider the lattice with nearest-neighbor interaction
that is described by (\ref{Main_Eq2}), where $K_{\alpha}(n-m)=0$, and 
\be \label{NNI}
\sum_{\substack{m=-\infty \\ m \ne n}}^{+\infty} \; K_2(n,m) \; u_m(t) =
u_{n+1}(t)-2u_n(t)+u_{n-1}(t) , 
\ee
where the term $K_2(n,m)$ describes the nearest-neighbor interaction.
Let us derive the usual elastic equation 
from the lattice model with the nearest-neighbor interaction
with coupling constant $g_2=K$, which is the spring stiffness. 
The following statement \cite{JMP2006,JPA2006,TarasovSpringer}  
gives for this lattice model with the nearest-neighbor interaction 
the corresponding continuum equation in the limit $d \to 0$. 

{\bf Proposition 1.} 
{\it In the continuous limit ($d \to 0$) the lattice equations of motion 
\be \label{CEM}
M \frac{d^2 u_n(t)}{d t^2} = 
K \cdot \Bigl( u_{n+1}(t)-2u_n(t)+u_{n-1}(t) \Bigr) + F (n) 
\ee
are transformed by the combination ${\cal F}^{-1} {\cal L} \ {\cal F}_{\Delta}$ 
of the operations (\ref{O1}-\ref{O3}) into the continuum equation: 
\be \label{CME0}
\frac{\partial^2 u(x,t)}{\partial t^2} = 
C^2_e \, \Delta u(x,t) + \frac{1}{\rho} f(x) ,
\ee
where 
\be \label{rEC}
\rho=\frac{M}{A \, d} , 
\qquad C^2_e =\frac{E}{\rho}=\frac{K \, d^2}{M} ,
\qquad E= \frac{K \, d}{A} , 
\ee
and $C^2_e$ is a finite parameter, $A$ is the cross-section area 
of the medium, $E$ is the Young's modulus, and 
$f(x) = F(x)/ (Ad)$ is the force density.} \\

A detailed proof of Proposition 1 is given in Appendix 1. \\

As a result, we prove that lattice equations (\ref{CEM}) 
in the limit $d \to 0$ give the continuum equation with the Laplacian (see also \cite{MPLB2014}).
Note that this result can be derived by methods described 
in Section 8 of \cite{Maslov}, where the relation
\[ \exp \, i \left( -i \, d \, \frac{\partial}{\partial x} \right) \, u(x,t) = u(x+d,t) \]
and the representation of (\ref{CEM}) by pseudo-differential equation are used.

%%%%%%%%%%%%%%%%%%%%%%%%%%%%%%%%%%%%%%%%%%%%%%%%%%%%%%%%%%%%%%%
%%%%%%%%%%%%%%%%%%%%%%%%%%%%%%%%%%%%%%%%%%%%%%%%%%%%%%%%%%%%%%%
%%%%%%%%%%%%%%%%%%%%%%%%%%%%%%%%%%%%%%%%%%%%%%%%%%%%%%%%%%%%%%%
\section{Lattice model with long-range interaction}

Let us derive a continuum equation for
the lattice with long-range interaction 
that is described by (\ref{Main_Eq2}), where $K_{\alpha}(n-m)$ satisfies the conditions
\be \label{Jnm}
K_{\alpha}(n-m)=K_{\alpha}(|n-m|) , \qquad \sum^{\infty}_{n=1} |K_{\alpha}(n)|^2 < \infty .
\ee
To have fractional gradient elasticity models, we assume that the function
\be \label{Jak}
\hat{K}_{\alpha}(k)=\sum^{+\infty}_{\substack{n=-\infty \\ n\not=0}} 
e^{-ikn} K_{\alpha}(n) = 2 \sum^{\infty}_{n=1} K_{\alpha}(n) \cos(kn) ,
\ee
satisfies the condition
\be \label{Aa}
\lim_{k \rightarrow 0} 
\frac{\hat{K}_{\alpha}(k) - \hat{K}_{\alpha}(0)}{|k|^{\alpha}} = A_{\alpha} ,
\ee
where $A_{\alpha}$ has a finite value. Condition (\ref{Aa}) means that 
\be \label{AR}
\hat{K}_{\alpha}(k)- \hat{K}_{\alpha}(0)= A_{\alpha} |k|^{\alpha} +R_{\alpha }(k),
\ee
for $k\rightarrow 0$, where
\be
 \lim_{k \rightarrow 0} \ R_{\alpha}(k) / |k|^{\alpha}  =0 .
\ee

The interaction terms $K_{\alpha}(|n-m|)$, which 
give the continuum equations of gradient elasticity models, 
can be defined as
\be \label{Kn}
K_{\alpha}(n) =\frac{(-1)^{n} }{\Gamma(\alpha/2+1+n) \Gamma(\alpha/2+1-n)} .
\ee
Using the series (Ref. \cite{Prudnikov}, Sec.5.4.8.12)
\be
\sum^{\infty}_{n=1}
\frac{(-1)^n}{\Gamma(\nu +1+n) \Gamma(\nu +1-n)} \cos(nk)
= \frac{2^{2 \nu -1}}{\Gamma(2 \nu +1)} 
\sin^{2\nu} \left(\frac{k}{2}\right) -\frac{1}{2\Gamma^2(\nu+1)} , \ee
where $\nu >-1/2$ and $0<k<2\pi$, we get
\be \label{alpha-1} 
\hat{K}_{\alpha }(k)-\hat{K}_{\alpha }(0)= \frac{2^{2 \alpha-1}}{\Gamma(\alpha+1)} 
\sin^{\alpha} \left(\frac{k}{2} \right) = \frac{1}{2\Gamma(\alpha+1)} \, k^{\alpha} + O(k^{\alpha +2}) . 
\ee
Here we use $\nu=\alpha /2$ and $ \sin (k/2) =k/2+ O(k^3)$. The limit $k \to 0$ gives
\be \label{kka}
\lim_{k \to 0} \frac{\hat{K}_{\alpha }(k)- \hat{K}_{\alpha }(0)}{|k|^{\alpha }} = 
\frac{1}{2 \Gamma(\alpha + 1)} ,
\ee
and we have $A_{\alpha} = 1 / (2 \Gamma (\alpha + 1))$. 
To consider a fractional generalization of the elastic theory, 
the variables $x$ and $d=\Delta x$ are dimensionless. 
Note that the interaction (\ref{Kn}) for integer 
values of $\alpha$ is discussed in \cite{MPLB2014}. \\

%%%%%%%%%%%%%%%%%%%%%%%%%%%%%%%%%%%%%%%%%%%%%%%%%%%%%%%%%%%%%

{\bf Proposition 2.}
{\it The lattice equations 
\be \label{C1}
M \frac{d^2 u_n}{d t^2} = 
g_2 \sum_{\substack{m=-\infty \\ m \ne n}}^{+\infty} \, K_2(n,m) \; \Bigl( u_n -u_m \Bigr) +
g_{\alpha} \sum_{\substack{m=-\infty \\ m \ne n}}^{+\infty} 
\, K_{\alpha} (n-m) \; \Bigl( u_n -u_m \Bigr) + F (n) ,
\ee 
where $g_2$ and $g_{\alpha}$ are coupling constants, 
$K_2(n,m)$ is defined by (\ref{NNI})
and $K_{\alpha}(|n-m|)$ is defined by (\ref{Kn}),
are transformed by the combination ${\cal F}^{-1} {\cal L} \ {\cal F}_{\Delta}$ 
of the operations (\ref{O1}-\ref{O3}) into the continuum equation: 
\be \label{CME}
\frac{\partial^2 u(x,t)}{\partial t^2} =
C_2 \, \Delta u(x,t) - 
C_{\alpha} \, (-\Delta)^{\alpha/2} u(x,t) + \frac{1}{\rho} f(x) ,
\ee
where $(-\Delta)^{\alpha/2}$ is the fractional Laplacian in the Riesz's form \cite{KST,SKM}, and
\be \label{G2G4}
C_2=\frac{g_2 \, d^2}{4 M} , \qquad C_{\alpha}= \frac{g_{\alpha} d^{\alpha}}{2\Gamma(\alpha+1) M} 
\ee
are finite parameters. } \\

A detailed proof of Proposition 2 is given in Appendix 2. \\

%%%%%%%%%%%%%%%%%%%%%%%%%%%%%%%%%%%%%%%%%%%%%%%%%%%%%%%%%%

In the Proposition 2, we use the Riesz fractional derivative 
$(- \Delta)^{\alpha/2}$. It can be defined as 
non-integer power of the Laplace operator 
in terms of the Fourier transform ${\cal F}$ by
\be \label{RFD-1}
((-\Delta)^{\alpha/2} f)(x)= {\cal F}^{-1} \Bigl( |k|^{\alpha} ({\cal F} f)(k) \Bigr) .
\ee
This fractional Laplacian can be also defined 
in the form of the hypersingular integral \cite{SKM,KST} by
\[ ((-\Delta)^{\alpha/2}f)(x) =\frac{1}{d_n(m,\alpha)} \int_{\mathbb{R}^n} 
\frac{1}{|z|^{\alpha+n}} (\Delta^m_z f)(z) \, dz , \] 
where $m> \alpha>0$, and $(\Delta^m_z f)(z)$ 
is the finite difference of order $m$ 
of a function $f(x)$ with a vector step $z \in \mathbb{R}^n$
and centered at the point $x \in \mathbb{R}^n$:
\[ (\Delta^m_z f)(z) =\sum^m_{k=0} (-1)^k \frac{m!}{k!(m-k)!}  \, f(x-kz) , \]
where the constant $d_n(m,\alpha)$ is defined by
\[ d_n(m,\alpha)=\frac{\pi^{1+n/2} A_m(\alpha)}{2^{\alpha} 
\Gamma(1+\alpha/2) \Gamma(n/2+\alpha/2) \sin (\pi \alpha/2)} ,  \]
and
\[ A_m(\alpha)=\sum^m_{j=0} (-1)^{j-1} \frac{m!}{j!(m-j)!} \, j^{\alpha} . \]
This hypersingular integral does not depend \cite{SKM,KST} 
on the choice of $m>\alpha>0$.

%%%%%%%%%%%%%%%%%%%%%%%%%%%%%%%%%%%%%%%%%%%%%%%%%%%%%%%%%%

We can note another possibility to set 
the interactions described by $K_2(n,m)$.
The term $K_2(n,m)$ that describes the nearest-neighbor interaction 
can be represented in forms that differ from (\ref{NNI}).
In general $K_2(n,m)$ describes a special form of the long-range interaction.
Let us give some example of these forms.

{\bf Example 1}. 
Instead of the nearest-neighbor interaction function (\ref{NNI}) 
we can use the long-range interaction with 
\be \label{2--1n2}
K_2(n,m)=\frac{(-1)^{|n-m|}}{|n-m|^2}  . 
\ee
Using (Ref. \cite{Prudnikov}, Sec. 5.4.2.12)
\[ \sum^{\infty}_{n=1} \frac{(-1)^n}{n^2} \cos (nk)=
\frac{1}{4}\left( k^2-\frac{\pi^2}{3} \right) , \quad 
|k| \leqslant \pi , \]
we obtain
\[ \hat{K}_2(k) =
2 \sum^{+\infty}_{n=1} \frac{(-1)^n}{n^2} \cos(kn)=
\frac{1}{2} k^2- \frac{\pi^2}{6} , \quad  |k| \leqslant \pi . \]
Then we have 
\be \label{2-Jk-1n2}
\hat{K}_2(k)-\hat{K}_2(0) = \frac{1}{2} \, k^2 .
\ee

{\bf Example 2}. 
If we consider the long-range interaction in the form
\be \label{example4}
K_2 (n)=\frac{(-1)^{n+1}}{n^2-a^2} , \ee
then equation (\ref{Jak}) gives
\[ \hat{K}(k)=\frac{\pi}{a\sin(\pi a)} \cos (ak) -\frac{1}{a^2} . \]
For $k \to 0$, we obtain
\be \label{2-54}
\hat{K}_2(k)- \hat{K}_2 (0) =
\frac{\pi a}{2 \sin(\pi a)} \, k^2 + O(k^4).
\ee

{\bf Example 3}. 
The long-range interaction 
\be \label{2-Jnnn}
K_2(n,m)=\frac{1}{|n-m|^{\alpha}} \ee
with the non-integer parameter $\alpha>3$,
gives (see Theorem 8.7 in \cite{TarasovSpringer}) the relation
\[ \hat{K}_2(k)-\hat{K}_2(0) = -\zeta(\alpha-2) \, k^2 +  . . . \  , \]
where $\zeta(z)$ is the Riemann zeta-function. \\

Proposition 2 allows to demonstrate the close relation 
between the lattice structure 
and the fractional gradient non-local continuum. 
Let us describe the well-known special cases.

Lattice equations (\ref{C1}) have 
two parameters $g_2$ and $g_{\alpha}$.
The corresponding continuum equation (\ref{CME})  
have two finite parameters $C_2$ and $C_{\alpha}$.
If we use
\[ g_2= 4 \, K, \qquad g_{\alpha}=0 . \]
then 
\[ C_2= C^2_e = K d^2/M, \quad C_{\alpha}=0 , \]
and equation (\ref{CME}) gives equation (\ref{CME0}). 
If we assume that  
\[ g_2=g_{\alpha}= 4 \, K ,   \]
then 
\be \label{52b}
C_2= C^2_e = \frac{K d^2}{M} , \quad 
C_{\alpha} = \frac{2 \, C^2_e \, d^{\alpha-2}}{\Gamma (\alpha+1)} \ee 
and we get relation 
\be \label{eq6b}
\frac{\partial^2 u(x,t)}{\partial t^2} = C^2_e \, 
\Delta u(x,t) - \frac{2 \, d^{\alpha-2} \, C^2_e}{\Gamma (\alpha+1)} \, 
(-\Delta)^{\alpha/2} u(x,t) + \frac{1}{\rho} f(x), 
\ee
where $C_e=\sqrt{E/\rho}$ is the elastic bar velocity. 
%%%%%%%%%%%%%%%%%%%%%%%%%%%%%%%%%%%%%%%%%%%%%%%%%%%%%%%%%%%%%%%
Let us give a remark about the scale parameter $l_s(\alpha)$.
Equation (\ref{52b}) can
lead to incorrect conclusion about the behavior 
of the scale parameter 
\be \label{la}
l^2_s(\alpha) = \frac{C_{\alpha}\, \rho}{E} =
\frac{C_{\alpha}}{C^2_e} \ee
for $d \to 0$ in the case $0<\alpha<2$. 
Using $C^2_e = K d^2 / M$, 
the parameter (\ref{la}) can be written as
\be \label{la2}
l^2_s(\alpha) =  
\frac{2 \, K \, d^{\alpha}}{\Gamma (\alpha+1)\, C^2_e \, M} .
\ee
Using that the value of $C^2_e$ is finite, then
behavior of the parameter $l^2_s(\alpha)$ for
$d \to 0$ has the same form for $\alpha>2$ and $0<\alpha<2$, 
such that  $l^2_s(\alpha)$ is proportion to $d^{\alpha}$.
Therefore we assume that the range of validity of alpha parameter is arbitrary real positive $\alpha$.

%%%%%%%%%%%%%%%%%%%%%%%%%%%%%%%%%%%%%%%%%%%%%%%%%%%%%%%%%%%%%%%

For $\alpha=4$ equation (\ref{eq6b}) is the usual equations of the gradient elasticity models
\be \label{eq6c}
\frac{\partial^2 u(x,t)}{\partial t^2} = C^2_e \, 
\Delta u(x,t) - \frac{d^{2} \, C^2_e}{12} \, 
\Delta^2 u(x,t) + \frac{1}{\rho} f(x) . 
\ee
The correspondent stress-strain relation for linear 
one-dimensional elasticity has the form
\[ \sigma(x,t) = E \Bigl(1 - l^2_s \Delta \Bigr) \varepsilon (x,t), \]
where $\sigma(x,t)$ is the stress, $\varepsilon (x,t)$
is the strain, and $l_s$ is the scale parameter.

In general, the coupling constants $g_2$ and $g_{\alpha}$ are independent. 
Therefore the sign of the coupling constant $g_{\alpha}$ (including the case $\alpha=4$) 
may differ from the sign of the constant $g_2=4 \, K$.
For $\alpha=4$ the second-gradient parameter is defined by the relation
\be
l^2_s = \frac{|g_4| \, d^2}{48 \, K} ,
\ee
where the sign in front of the factor $l^2_s$ 
is determined by the sign of the coupling constant $g_4$.
If the constant $g_4$ is positive then we get 
the gradient elasticity model with negative sign \cite{AA2011}.

As a result the second-gradient model with negative and positive sign 
can be derived from a microstructure of lattice particles by suggested approach.
The suggested approach as shown above uniquely leads to second-order
strain gradient terms that are preceded by the positive and negative signs.

%%%%%%%%%%%%%%%%%%%%%%%%%%%%%%%%%%%%%%%%%%%%%%%%%%%%%%%%%%%%%%

\section{Stationary Solution for Fractional Gradient Elasticity}

%%%%%%%%%%%%%%%%%%%%%%%%%%%%%%%%%%%%%%%%%%%%%%%%%%%%%%%%%%%%%%

We can consider more general model of lattice with long-range interaction, 
where all particles are displaced from its equilibrium in one direction, 
and the displacement of particles is described by a scalar field $u({\bf r},t)$, where ${\bf r} \in \mathbb{R}^n$ ($n=1,2,3$).
The correspondent continuum equation of 
the fractional elasticity model is
\be \label{1-CME-3}
\frac{\partial^2 u({\bf r},t)}{\partial t^2} =
C_2 \, \Delta u({\bf r},t) -
C_{\alpha} \, (-\Delta)^{\alpha/2} u ({\bf r},t) + \frac{1}{\rho} f({\bf r}) ,
\ee
where ${\bf r}$ and $r=|{\bf r}|$ are dimensionless variables. 

Let us consider the static case ($\partial u({\bf r},t) /\partial t =0$, i.e. $u({\bf r},t) = u({\bf r})$) 
in this fractional elasticity model.
Then equation (\ref{1-CME-3}) has the form
\be \label{FPDE-1}
C_2 \, \Delta u({\bf r}) - C_{\alpha} \, (-\Delta)^{\alpha/2} u({\bf r}) 
+ \frac{1}{\rho} f({\bf r}) = 0  .
\ee

We can use the Fourier method to solve 
fractional differential equation (\ref{FPDE-1}), 
which is based on the relations 
\be \label{FFL2}
{\cal F}[ (-\Delta)^{\alpha/2} u({\bf r})]({\bf k})= 
|{\bf k}|^{\alpha} \, \hat u({\bf k}), \quad
{\cal F}[ \Delta u({\bf r})]({\bf k})= 
- \, {\bf k}^2 \, \hat u({\bf k}).
\ee
Applying the Fourier transform ${\cal F}$ to both sides of (\ref{FPDE-1}) and using (\ref{FFL2}), we have
\be
({\cal F} u)({\bf k}) = \frac{1}{\rho} \,
\left( C_2 \, |{\bf k}|^2 + C_{\alpha} \, |{\bf k}|^{\alpha_j} \right)^{-1} \, ({\cal F} f)({\bf k}) .
\ee

Equation (\ref{FPDE-1}) (see, for example, Section 5.5.1. in \cite{KST}) 
has a particular solution that 
can be represented in the form of the convolution of the 
functions $G^n_{\alpha}(|{\bf r}|)$ and $f(|{\bf r}|)$ as follow
\be \label{phi-G}
u ({\bf r})=  \frac{1}{\rho} \, \int_{\mathbb{R}^n} G^n_{\alpha} ({\bf r} - {\bf r}^{\prime}) \, 
f ({\bf r}^{\prime}) \, d^n {\bf r}^{\prime} ,
\ee
where $G^n_{\alpha}({\bf r})$  is the Green function (see Section 5.5.1. in \cite{KST}) 
of the form
\be \label{FGF}
G^n_{\alpha}({\bf r})= {\cal F}^{-1} 
\Bigl[ \left( C_2 \, |{\bf k}|^2 + C_{\alpha} \, |{\bf k}|^{\alpha} \right)^{-1} \Bigr] ({\bf r})=
\int_{\mathbb{R}^n} \left( C_2 \, |{\bf k}|^2 + C_{\alpha} \, |{\bf k}|^{\alpha} \right)^{-1} \
e^{ + i ({\bf k},{\bf r}) } \, d^n {\bf k} .
\ee

We can use the relation
\be \label{3-1}
\int_{\mathbb{R}^n} e^{  i ({\bf k},{\bf r}) } \, f(|{\bf k}|) \, d^n {\bf k}= 
\frac{(2 \pi)^{n/2}}{ |{\bf r}|^{(n-2)/2}} 
\int^{\infty}_0 f( \lambda) \, \lambda^{n/2} \, J_{n/2-1}(\lambda |{\bf r}|) \, d \lambda
\ee
that holds (see Lemma 25.1 of \cite{SKM}) for any suitable function $f$
such that the integral in the right-hand side of (\ref{3-1}) is convergent. 
Here $J_{\nu}$ is the Bessel function of the first kind. 

%%%%%%%%%%%%%%%%%%%%%%%%%%%%%%%%%%%%%%%%%%%%%%%%%%%%%%%%%%%%%%

Using relation (\ref{3-1}), the Green function (\ref{FGF}) 
can be represented (see Theorem 5.22 in \cite{KST})
in the form of the integral with respect to one parameter $\lambda$ as
\be \label{G-1}
G^n_{\alpha} ({\bf r}) = \frac{|{\bf r}|^{(2-n)/2}}{(2 \pi)^{n/2}} 
\int^{\infty}_0 \frac{\lambda^{n/2} \, J_{(n-2)/2} (\lambda |{\bf r}|) \, d \lambda}{
C_2 \, \lambda^2 + C_{\alpha} \, \lambda^{\alpha} } ,
\ee
where $n=1,2,3$, and 
$J_{(n-2)/2}$ is the Bessel function of the first kind.

For the 3-dimensional case ($n=3$), we can use 
\be
J_{1/2} (z) = \sqrt{\frac{2}{\pi z}} \, \sin (z),
\ee
and we have
\be \label{G-1-3D}
G^3_{\alpha} ({\bf r}) =\frac{1}{ 2 \pi^2 |{\bf r}| } 
\int^{\infty}_0 \frac{\lambda \, \sin (\lambda |{\bf r}|) \, d \lambda}{
C_2 \, \lambda^2 + C_{\alpha} \, \lambda^{\alpha} }  .
\ee
For the 1-dimensional case ($n=1$), we can use 
\be
J_{-1/2} (z) = \sqrt{\frac{2}{\pi z}} \, \cos (z) ,
\ee
and we have (see Theorem 5.24 in \cite{KST}) the function
\be 
G^1_{\alpha} ({\bf r}) =\frac{1}{\pi} 
\int^{\infty}_0 \frac{\cos (\lambda |{\bf r}|) \, d \lambda}{
 C_2 \, \lambda^2 + C_{\alpha} \, \lambda^{\alpha} }  .
\ee

%%%%%%%%%%%%%%%%%%%%%%%%%%%%%%%%%%%%%%%%%%%%%%%%%%%%%%%%%%%%%%

Let us determine the deformation of an infinite 
elastic continuum,
when a force is applied to a small region of the medium.
This is the well-known Thomson's problem \cite{LL}.
We solve Thomson's problem in the framework of  
the fractional elasticity model. 
If we consider the deformation for $|{\bf r}|$, 
which are larger compare with the size of the region,
we can suppose that the force is applied at a point. 
In this case, we have
\be \label{deltaf}
f({\bf r}) = f_0 \, \delta ({\bf r}) = f_0 \, \delta (x) \delta (y) \delta (z)  . 
\ee
Then the displacement field $u ({\bf r})$ of 
fractional elasticity  has a simple form of 
the particular solution  
that is proportional to the Green function
\be \label{phi-Gb}
u ({\bf r}) = \frac{f_0}{\rho} \, G^n_{\alpha} ({\bf r}) .
\ee
Therefore, the displacement field for the case (\ref{deltaf}) 
has the form
\be \label{Pot-2}
u ({\bf r}) = \frac{1}{2 \pi^2} \frac{f_0}{ \rho \, |{\bf r}|} \, 
\int^{\infty}_0 \frac{ \lambda \, \sin (\lambda |{\bf r}|)}{ 
C_2 \, \lambda^2 + C_{\alpha} \, \lambda^{\alpha}  } \, d \lambda .
\ee
%%%For one-dimensional case, we have
%%%\be \label{Pot-4} u ({\bf r}) = \frac{f_0}{ \pi \, \rho } \, 
%%%\int^{\infty}_0 \frac{ \cos (\lambda |{\bf r}|)}{ 
%%%C_2 \, \lambda^2 + C_{\alpha} \, \lambda^{\alpha}  } \, d \lambda . \ee

The asymptotic behavior $|{\bf r}| \to \infty$ of the displacement field 
$u ({\bf r})$  in the model described by (\ref{Pot-2}) with (\ref{deltaf}),
is given by 
\be \label{As-11}
u ({\bf r}) \ \approx \  \frac{f_0 \, \Gamma(2-\alpha) \, \sin(\pi \alpha/2)}{ \pi^3 \, C_{\alpha} \, \rho} \,
\cdot \,  \frac{1}{|{\bf r}|^{3-\alpha}} \quad (\alpha< 2) ,
\ee
\be \label{As-12}
u ({\bf r}) \ \approx \ \frac{1}{2 \pi^2} \frac{f_0}{ \rho \, |{\bf r}|} , \quad (\alpha>2) . 
\ee
Note that the asymptotic behavior $ |{\bf r}| \to \infty$ does not depend on the parameter $\alpha$ for $\alpha >2$.
In the case $\alpha<2$ the displacement field on the long distances is determined only by term 
with the fractional Laplacian of the order $\alpha$.

The asymptotic behavior $|{\bf r}| \to 0$ of the displacement field 
$u ({\bf r})$  that is described by equation (\ref{Pot-2}), 
where the force $f({\bf r})$ is applied 
at a point (\ref{deltaf}), is given by 
\be \label{As-21}
u ({\bf r}) \ \approx \ \frac{1}{2 \pi^2} \frac{f_0}{ \rho \, |{\bf r}|} , \quad (\alpha<2),
\ee
\be \label{As-22}
u ({\bf r}) \ \approx \ 
\frac{f_0 \, \Gamma((3-\alpha)/2)}{ 2^{\alpha} \, \pi^2 \sqrt{\pi} \, \rho \, C_{\alpha} \, \Gamma(\alpha/2)} \,
\cdot \, \frac{1}{|{\bf r}|^{3-\alpha}} , \quad (2<\alpha<3),
\ee
\be \label{As-23}
u ({\bf r}) \ \approx \ 
\frac{f_0}{2 \pi \, \alpha \, \rho \, c^{1-3/\alpha}_{\beta} \, c^{3/\alpha}_{\alpha} \, \sin (3 \pi / \alpha)}
 , \quad (\alpha>3) .
\ee
Here the Euler's reflection formula for Gamma function is used.
Note that the asymptotic behavior $ |{\bf r}| \to 0$ does not depend on the parameter $\alpha$ for $\alpha <2$.
In the case $\alpha>2$, the displacement field on the 
short distances is determined only by term 
with the fractional Laplacian of the order $\alpha$.

The functions
\[  u (x) = \frac{1}{x} \, \int^{\infty}_0 \frac{ \lambda \, \sin (\lambda x)}{ 
C_2 \, \lambda^2 + C_{\alpha} \, \lambda^{\alpha}  } \, d \lambda \]
for the different orders of $1 < \alpha < 6$ and with $C_2=C_{\alpha}=1$ are present on Figures 1-4, where $x=|{\bf r}|$.
Figures 1,2,4 allows us to see that 
the field  $u (x)$  tends to a constant value
($u (x) \to const$) at $x \to 0$ for the parameters
$\alpha > 3$ ($\alpha=3.6$, $\alpha=4.1$, $\alpha=5.2$ and $\alpha=5.6$).
Figures 2 and 3 demonstrate that 
the asymptotic behavior of the type 
$u (x) \ \approx \ 1 / x^{3-\alpha}$ 
for the field  $u (x)$ 
at $x \to 0$ for the parameters $2<\alpha<3$ 
($\alpha=2.6$, $\alpha=2.7$).
Figures 3 and 4 show that 
the asymptotic behavior of the type 
$u (x) \ \approx \ 1 / x$ 
for the field  $u (x)$ 
at $x \to 0$ for the parameters $0<\alpha<2$ 
($\alpha=1.4$, $\alpha=1.9$).

%%%%%%%%%%%%%%%%%%%%%%%%%%%%%%%%%%%%%%%%%%%%%%%%%%%%%%%%%%%%%

We can determine the deformation of an 
infinite non-local elastic continuum, when 
a pair of forces with equal in magnitude and oppositely directed
is applied to a small region of the medium.
We assume that these forces are separated by small distance. 
If we consider the deformation for $|{\bf r}|$, 
which are larger compare with the size of the region,
we can suppose that two force are applied at two points such that 
\be \label{deltaf2}
f({\bf r}) = f_0 \, \delta ({\bf r}+{\bf a}) - f_0 \, \delta ({\bf r}-{\bf a})  . 
\ee
This problem is analogous to a dipole system as 
a pair of electric charges of equal magnitude but opposite sign, 
separated by small distance.
Then the displacement field $u ({\bf r})$ of 
fractional elasticity  has a form of 
the particular solution 
\be \label{phi-Gb2}
u ({\bf r}) = \frac{f_0}{\rho} \, 
\Bigl( G^n_{\alpha} ({\bf r}+{\bf a}) - G^n_{\alpha} ({\bf r}-{\bf a}) \Bigr).
\ee
Therefore, the displacement field for the case (\ref{deltaf2}) 
has the form
\be \label{Pot-2-Dip}
u ({\bf r}) = \frac{1}{2 \pi^2} \frac{f_0}{ \rho } \, 
\int^{\infty}_0 \Bigl( \frac{ \sin (\lambda |{\bf r}+{\bf a}| )}{ |{\bf r}+{\bf a}| } 
\frac{ \sin (\lambda |{\bf r}-{\bf a}|)}{ |{\bf r}-{\bf a}| }  \Bigr)
\frac{ \lambda }{ C_2 \, \lambda^2 + C_{\alpha} \, \lambda^{\alpha}  } 
\, d \lambda .
\ee
For $|{\bf r}| \gg |{\bf a}|$, we can use
$|{\bf r}-{\bf a}| - |{\bf r}+{\bf a}| \approx 2 \, |{\bf a}| \, \cos \theta$,
where $\theta$ is the angle between ${\bf r}$ and ${\bf a}$, 
and 
$|{\bf r}-{\bf a}| \, |{\bf r}+{\bf a}| \approx |{\bf r}|^2$.
Then the displacement field can be represented in the form
\be \label{Pot-2-Dip2}
u ({\bf r}) = \frac{ f_0 \, |{\bf a}| \, \cos \theta}{\pi^2 \, \rho \, |{\bf r}|^2} \, 
\int^{\infty}_0 \frac{ \lambda \, \sin (\lambda |{\bf r}|)}{ 
C_2 \, \lambda^2 + C_{\alpha} \, \lambda^{\alpha}  } \, d \lambda
+
\frac{ f_0 }{\pi^2 \, \rho \, |{\bf r}|} \, 
\int^{\infty}_0 \frac{ \lambda \, \, \sin( \lambda \, |{\bf a}| \, \cos \theta ) \, \cos (\lambda |{\bf r}|)}{ 
C_2 \, \lambda^2 + C_{\alpha} \, \lambda^{\alpha}  } \, d \lambda
\ee

%%%%%%%%%%%%%%%%%%%%%%%%%%%%%%%%%%%%%%%%%%%%%%%%%%%%%%%%%%%%%

For the suggested lattice equation and correspondent continuum limit, we can use all positive values of alpha parameter. There is no reason to limit of the range alpha values in the used fractional differential equations for the suggested form of fractional gradient and integral elasticity. It allows us to state that the range of validity of alpha parameter is arbitrary real positive values. As a result,
we can distinguish two following particular cases
in the fractional elasticity model described by (\ref{FPDE-1}):
(1) fractional integral elasticity ($\alpha<2$); 
(2) fractional gradient elasticity ($\alpha > 2$). 
Note that for the first case the order of the fractional Laplacian in equation (\ref{1-CME-3}) is
less than the order of the term related to the usual Hooke's law. 
In the second case the order of the fractional Laplacian is greater 
of the order of the term related to the Hooke's law.

\newpage
%%%%%%%%%%%%%%%%%%%%%%%%%%%%%%%%%%%%%%%%%%%%%%%%%%%%%%%%%%

\section{Conclusion}

We suggest lattice models with long-range interaction
of power-law type
as microscopic model of fractional non-local elastic continuum. 
The continuum equations with derivatives of non-integer orders
are directly derived from the suggested lattice models. 
We prove that the fractional gradient and fractional 
integral models can be derived 
from lattice models 
with long-range particle interactions.
The suggested approach uniquely leads 
to second-order and fractional-order strain gradient terms 
that are preceded by the positive and negative signs.
Fractional calculus 
allows us to obtain exact analytical solutions of
the fractional differential equations 
for continuum models of a wide class of material 
with fractional gradient and fractional integral non-locality.
%%%The fractional equation allows us to describe
%%%the dynamics of the complex non-local elastic materials.
A characteristic feature of the behavior 
of a fractional non-local continuum 
is the spatial power-tails of non-integer orders 
in the asymptotic behavior.
The fractional elasticity models, 
which are suggested in this paper to describe 
complex materials with fractional non-locality, 
can be characterized 
by a common or universal spatial behavior of elastic materials 
by analogy with the universal temporal behavior 
of low-loss dielectrics \cite{Jo1,Jo2,Jo3,JPCM2008-1}.
The asymptotic behavior (\ref{As-11}) and (\ref{As-21}) 
allows us to state that fractional integral elasticity effects
are important on the macroscopic scales.
The asymptotic behavior (\ref{As-12}) and (\ref{As-22}-\ref{As-23}) 
allows us to state that fractional gradient elasticity effects
are very important for the mesoscopic and nano scales.
As a results the fractional gradient elasticity models
can be very important for nanomechanics 
\cite{Nano-1,Nano-2,Nano-3,Nano-4,Nano-5}
of nonlocal materials with long-range particle interactions.

%%%%%%%%%%%%%%%%%%%%%%%%%%%%%%%%%%%%%%%%%%%%%%%%%%%%%%%%%%%%%%%%%%%%%%%%%

%%% ------------------------ PLOT --------------------------

%%%\newpage
%%%%%%%%%%%%%%%%%%%%%%%%%%%%%%%%%%%%%%%%%%%%%%%%%%%%%%%%%%%%%%%

%%%\newpage
%%%%%%%%%%%%%%%%%%%%%%%%%%%%%%%%%%%%%%%%%%%%%%%%%%%%%%%%%%%%%%
\begin{figure}[H]
\begin{minipage}[h]{0.47\linewidth}
\resizebox{11cm}{!}{\includegraphics[angle=-90]{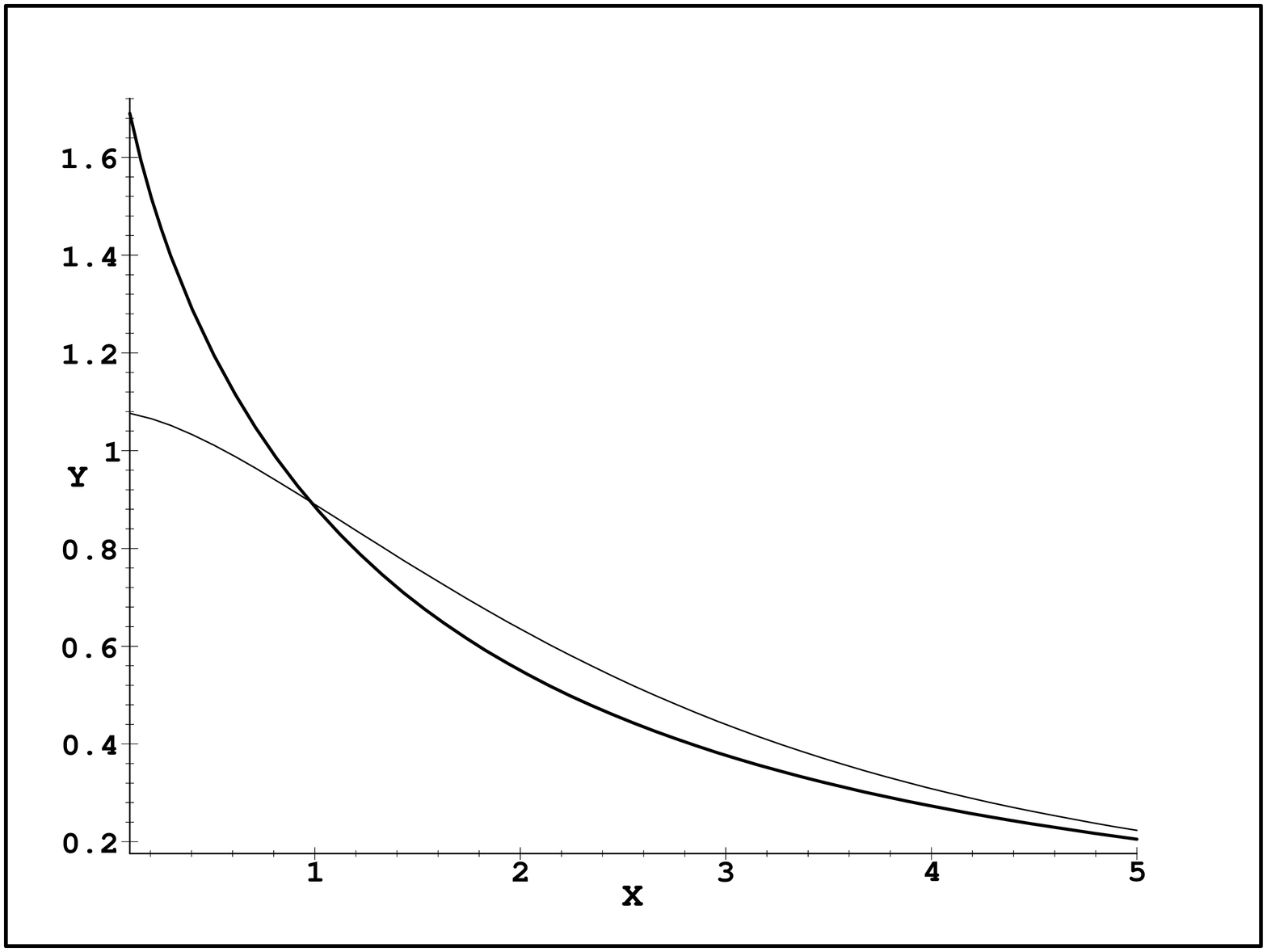}} 
\end{minipage}
\caption{Plots of function $u(x)$ for the orders  
$\alpha=3.6$ and $\alpha=5.2$,
where $x=|{\bf r}|$ and $C_2=C_{\alpha}=1$.}
\label{Plot1}
\end{figure}

\begin{figure}[H]
\begin{minipage}[h]{0.47\linewidth}
\resizebox{11cm}{!}{\includegraphics[angle=-90]{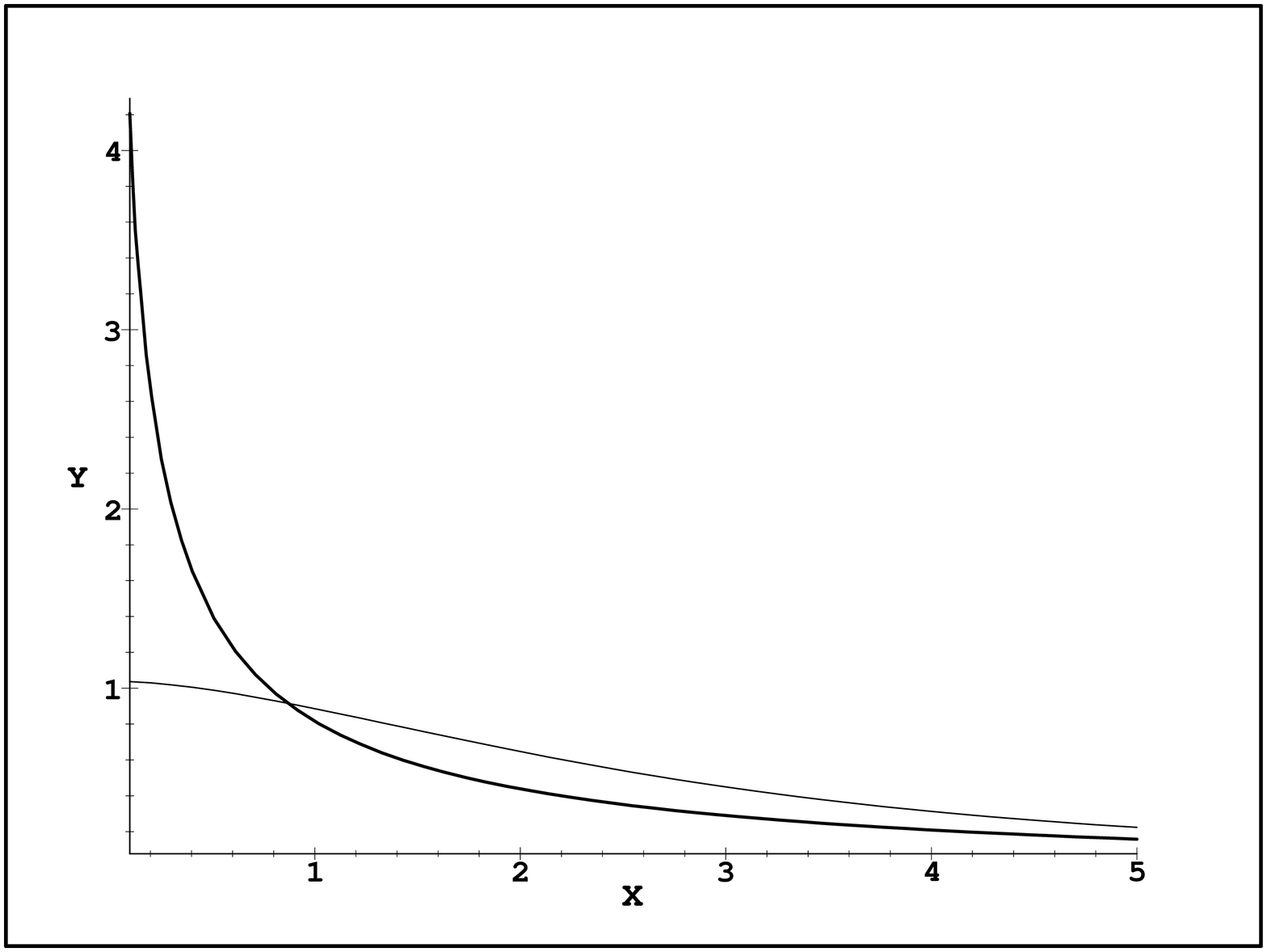}} 
\end{minipage}
\caption{Plots of function $u(x)$ for the orders  
$\alpha=2.6$ and $\alpha=5.6$, 
where $x=|{\bf r}|$ and $C_2=C_{\alpha}=1$.}
\label{Plot2}
\end{figure}

\begin{figure}[H]
\begin{minipage}[h]{0.47\linewidth}
\resizebox{11cm}{!}{\includegraphics[angle=-90]{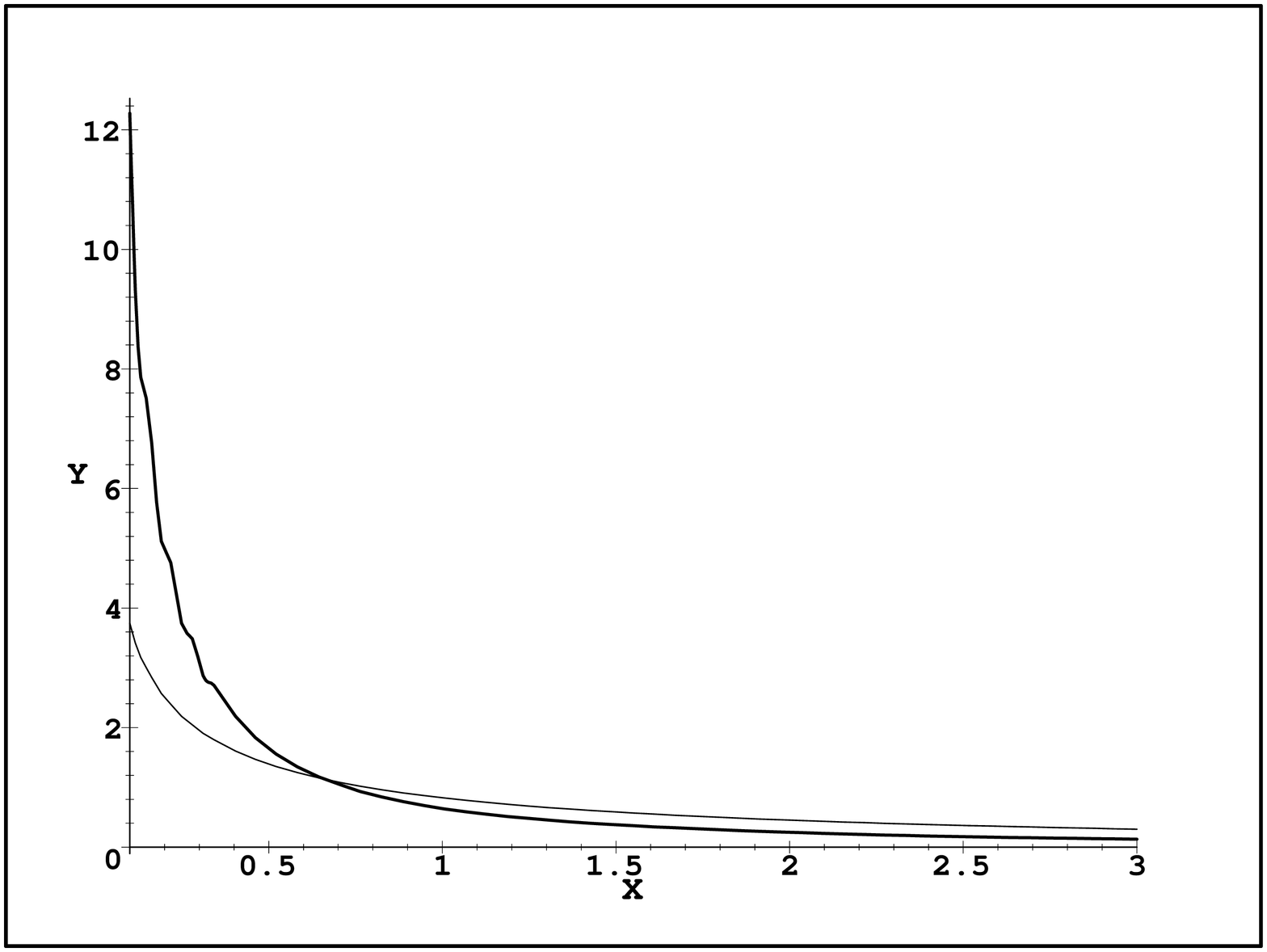}} 
\end{minipage}
\caption{Plots of function $u(x)$ for the orders  
$\alpha=1.4$ and $\alpha=2.7$,
where $x=|{\bf r}|$ and $C_2=C_{\alpha}=1$.}
\label{Plot3}
\end{figure}

\begin{figure}[H]
\begin{minipage}[h]{0.47\linewidth}
\resizebox{11cm}{!}{\includegraphics[angle=-90]{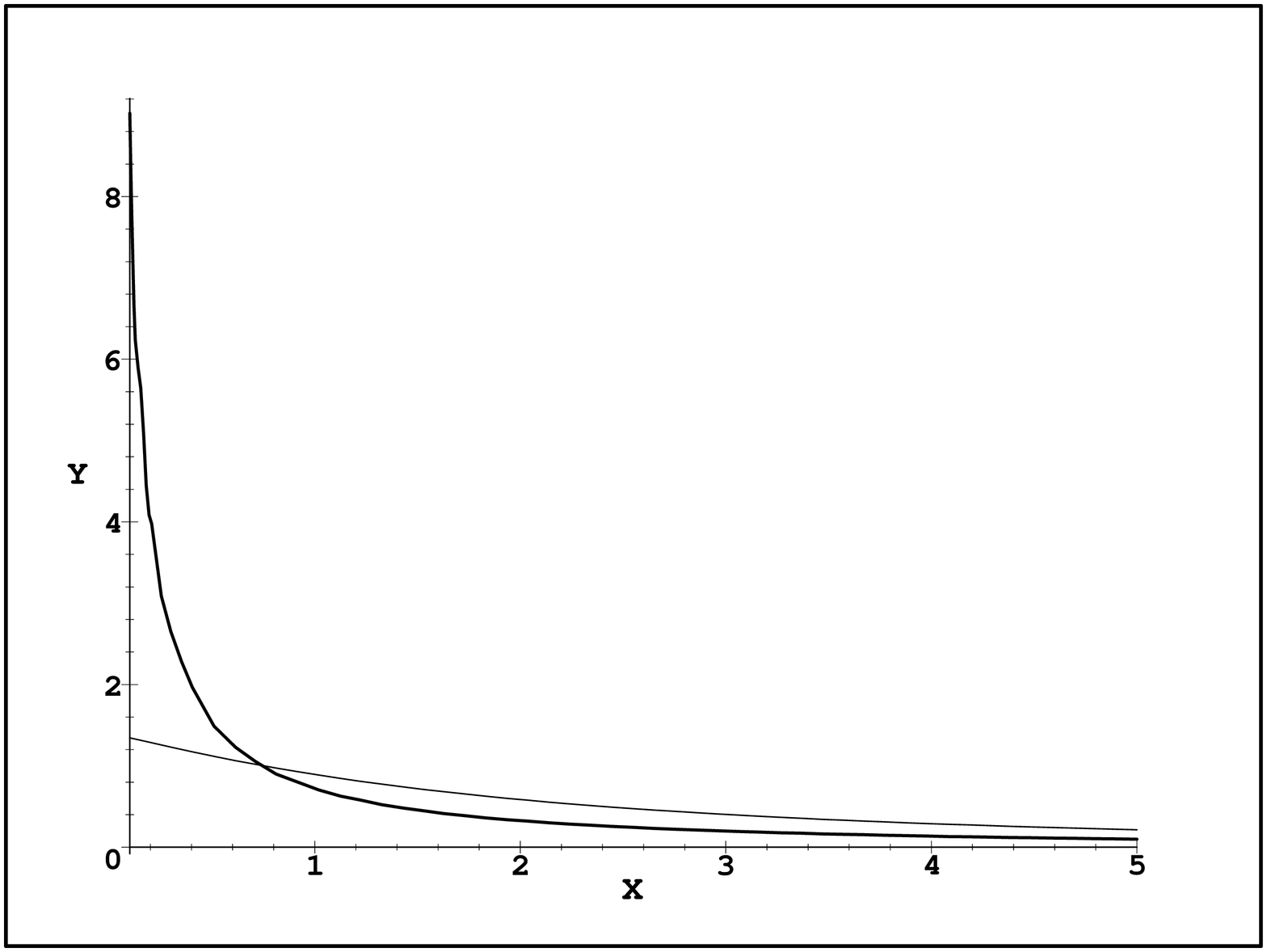}} 
\end{minipage}
\caption{Plots of function $u(x)$ for the orders 
$\alpha=1.9$ and $\alpha=4.1$,
where $x=|{\bf r}|$ and $C_2=C_{\alpha}=1$.}
\label{Plot4}
\end{figure}

%%%%%%%%%%%%%%%%%%%%%%%%%%%%%%%%%%%%%%%%%%%%%%%%%%%%%%%%%%%%%%%%%
%%%%%%%%%%%%%%%%%%%%%%%%%%%%%%%%%%%%%%%%%%%%%%%%%%%%%%%%%%%%%%%%%
%%%%%%%%%%%%%%%%%%%%%%%%%%%%%%%%%%%%%%%%%%%%%%%%%%%%%%%%%%%%%%%%%
%%% \newpage

%%%%%%%%%%%%%%%%%%%%%%%%%%%%%%%%%%%%%%%%%%%%%%%%%%%%%%%%%%%%%%
%%%%%%%%%%%%%%%%%%%%%%%%%%%%%%%%%%%%%%%%%%%%%%%%%%%%%%%%%%%%%%
%%%%%%%%%%%%%%%%%%%%%%%%%%%%%%%%%%%%%%%%%%%%%%%%%%%%%%%%%%%%%%

\section*{Appendix 1: Proof of Proposition 1}

To derive the equation for the field $\hat u(k,t)$, we
multiply equation (\ref{CEM}) by $\exp(-ikn d )$, 
and summing over $n$ from $-\infty$ to $+\infty$. Then
\be \label{DD1}
\sum^{+\infty}_{n=-\infty} e^{-ikn d } M \, \frac{d^2 u_n}{d t^2} =
K \cdot \sum^{+\infty}_{n=-\infty} \, e^{-ikn d } \, \Bigl( u_{n+1}-2u_n+u_{n-1} \Bigr) +
\sum^{+\infty}_{n=-\infty} e^{-iknd } F(n) .
\ee
The first term on the right-hand side of (\ref{DD1}) is
\[
K \cdot \sum^{+\infty}_{n=-\infty} \
e^{-ikn d } K_2(n,m) u_n = K \cdot \sum^{+\infty}_{n=-\infty} \
e^{-ikn d } \, \Bigl( u_{n+1}-2u_n+u_{n-1} \Bigr) 
= \]
\[ = K \cdot \sum^{+\infty}_{n=-\infty} \
e^{-ikn d }  u_{n+1} -
2 K \cdot \sum^{+\infty}_{n=-\infty} \
e^{-ikn d }  u_n +
K \cdot \sum^{+\infty}_{n=-\infty} \
e^{-ikn d }  u_{n-1}= \]
\[ =e^{ikd } \, 
K \cdot \sum^{+\infty}_{m=-\infty} \
e^{-ik m d }  u_{m} -
2 K \cdot \sum^{+\infty}_{n=-\infty} \
e^{-ikn d }  u_n + e^{-ik d } 
K \cdot \sum^{+\infty}_{j=-\infty} \
e^{-ik j d }  u_{j} .
\]
Using the definition of $\hat{u}(k,t)$, we obtain
\[ K \cdot \sum^{+\infty}_{n=-\infty} \
e^{-ikn d } K_2(n,m) u_n=
K \cdot \Bigl( e^{ikd } \hat{u}(k,t)- 2 \hat{u}(k,t) +
e^{-ik d } \hat{u}(k,t) \Bigr)= \]
\[ = K \cdot \Bigl( e^{ikd } +e^{-ik d }-2 \Bigr) \hat{u}(k,t)= 
2 K \cdot \Bigl( \cos \left( k d  \right)-1 \Bigr) \, \hat{u}(k,t) . \]
As a result, we have
\be \label{Proof-nearest1}
K \cdot \sum^{+\infty}_{n=-\infty} \
e^{-ikn d} K_2(n,m) u_n=
-4 K \cdot \sin^2 \left( \frac{kd}{2}  \right) \hat{u}(k,t) .
\ee
Substitution of (\ref{Proof-nearest1}) into (\ref{DD1}) gives
\be \label{simple}
M \frac{\partial^2 \hat{u}(k,t)}{\partial t^2} = 
-4 K \cdot \sin^2 \left( \frac{kd}{2}  \right) \hat{u}(k,t) +
\mathcal{F}_{\Delta} \{ F \left( u_n(t) \right) \} .
\ee
For $d  \to 0$, the asymptotic behavior of the sine is
$\sin( k d /2 ) = k d/2 + O((kd)^3) $. 
Then 
\[  - 4 \, \sin^2 \left( \frac{kd}{2}  \right) 
= - \left( k d \right)^2 + O((kd)^4) . \]
Using the finite parameter $C^2_e=K \, d^2/M$, 
the transition to the limit $d  \to 0$ 
in equation (\ref{simple}) gives 
\be \label{DD2}
\frac{\partial^2  \tilde u(k,t)}{\partial t^2}=
- C^2_e k^2 \tilde u(k,t) + \frac{1}{M} {\cal F} \{F(x)\} ,
\ee
where $C^2_e$ is defined by (\ref{rEC}).
The inverse Fourier transform ${\cal F}^{-1}$ of (\ref{DD2}) has the form
\be \frac{\partial^2 {\cal F}^{-1}\{ \tilde u(k,t)\} }{\partial t^2}=
- C^2_e {\cal F}^{-1} \{k^2 \tilde u(k,t)\} + \frac{1}{\rho} f(x) , \ee
where $f(x)=F(x)/(A\, d)$ is the force density,
and $\rho=M/(A\, d)$ is the mass density.
Then using 
\[ {\cal F}^{-1}\{ \tilde u(k,t)\} = u(x,t) , \quad
{\cal F}^{-1} \{k^2 \tilde u(k,t)\} = -\Delta u(x,t) , \]
we obtain the continuum equation (\ref{CME0}). This ends the proof. 

%%%%%%%%%%%%%%%%%%%%%%%%%%%%%%%%%%%%%%%%%%%%%%%%%%%%%%%%%%%%%%%%

\section*{Appendix 2: Proof of Proposition 2}

To derive the equation for the field $\hat u(k,t)$, we
multiply equation (\ref{C1}) by $\exp(-ikn d )$, 
and summing over $n$ from $-\infty$ to $+\infty$. Then
\[
\sum^{+\infty}_{n=-\infty} e^{-ikn d } M \,
\frac{d^2}{d t^2}u_n(t)=  g_2 \, \sum^{+\infty}_{n=-\infty} \
\sum^{+\infty}_{\substack{m=-\infty \\ m \not=n}}
e^{-ikn d }  K_2(n,m) \ \Bigl( u_n-u_m \Bigr) +
\]
\be \label{C3a}
+ g_{\alpha} \, \sum^{+\infty}_{n=-\infty} \
\sum^{+\infty}_{\substack{m=-\infty \\ m \not=n}}
e^{-ikn d }  K_{\alpha}(n-m) \ \Bigl( u_n-u_m \Bigr) +
\sum^{+\infty}_{n=-\infty} e^{-iknd } F(n) .
\ee
The left-hand side of (\ref{C3a}) gives
\be
\sum^{+\infty}_{n=-\infty} e^{-ikn d } 
\frac{\partial^2 u_n(t)}{\partial t^2}=
\frac{\partial^2 }{\partial t^2}
\sum^{+\infty}_{n=-\infty} e^{-ikn d } u_n(t)=
\frac{\partial^2 \hat{u}(k,t)}{\partial t^2} ,
\ee
where $\hat{u}(k,t)$ is defined by (\ref{ukt}).
The second term of the right-hand side of (\ref{C3a}) is
\be
\sum^{+\infty}_{n=-\infty} e^{-ikn d } F(n)=
{\cal F}_{\Delta} \{F(n)\} .
\ee

The limit for the first term on the right-hand side of (\ref{C3a}) is
described in Proposition 1.

The second term on the right-hand side of (\ref{C3a}) 
with a multiplier $g_{\alpha}$ is
\[
\sum^{+\infty}_{n=-\infty} \ \sum^{+\infty}_{\substack{m=-\infty \\ m \not=n}} 
e^{-ikn d } K_{\alpha}(n-m) \Bigl( u_n-u_m \Bigr) = \]
\be \label{C6}
=\sum^{+\infty}_{n=-\infty} \  \sum^{+\infty}_{\substack{m=-\infty \\ m \not=n}}
e^{-ikn d } K_{\alpha}(n-m) u_n - 
\sum^{+\infty}_{n=-\infty} \sum^{+\infty}_{\substack{m=-\infty \\ m \not=n}} 
e^{-ikn d } K_{\alpha}(n-m) u_m .
\ee
Using (\ref{ukt}), the first term in the right-hand side of (\ref{C6}) gives
\be \label{C7} 
\sum^{+\infty}_{n=-\infty} \ \sum^{+\infty}_{\substack{m=-\infty \\ m \not=n}}
e^{-ikn d } K_{\alpha}(n-m) u_n =
\sum^{+\infty}_{n=-\infty} e^{-ikn d } u_n 
\sum^{+\infty}_{\substack{m^{\prime}=-\infty \\ m^{\prime} \not=0}}
K_{\alpha}(m^{\prime})= \hat{K}_{\alpha} (0) \, \hat u(k,t) ,
\ee
where we use (\ref{Jnm}) and $K_{\alpha}(m^{\prime}+n-n)=K_{\alpha}(m^{\prime})$, and
\be \label{not}
\hat{K}_{\alpha}(k d )=
\sum^{+\infty}_{\substack{n=-\infty \\ n\not=0}} 
e^{-ikn d } K_{\alpha}(n)={\cal F}_{\Delta}\{ K_{\alpha}(n)\} .
\ee
Note that
\[ \sum^{+\infty}_{n=-\infty} \ 
\sum^{+\infty}_{\substack{m=-\infty \\ m \not=n}}
e^{-ikn d } K_{\alpha}(n-m) u_m = 
\sum^{+\infty}_{\substack{n=-\infty \\ n \not=m}} 
e^{-ikn d } K_{\alpha}(n-m) \ \sum^{+\infty}_{m=-\infty} u_m = \]
\be \label{C9}
= \sum^{+\infty}_{\substack{n^{\prime}=-\infty \\ n^{\prime}\not=0}} 
e^{-ikn^{\prime} d } K_{\alpha}(n^{\prime}) \
\sum^{+\infty}_{m=-\infty } u_m e^{-ikm d }=
\hat{K}_{\alpha} (k d ) \ \hat u(k,t) ,
\ee
where $K_{\alpha}(m-(n^{\prime}+m))=K_{\alpha}(n^{\prime})$ is used.

As a result, equation (\ref{C3a}) has the form
\be \label{20}
M \frac{\partial^2  \hat u(k,t)}{\partial t^2}= 
g_2\, \Bigl( \hat{K}_2 (0)- \hat{K}_2 (k d ) \Bigr) \, \hat u(k,t) 
+ g_{\alpha}\, 
\Bigl( \hat{K}_{\alpha} (0)- \hat{K}_{\alpha} (k d ) \Bigr) \, \hat u(k,t) 
+{\cal F}_{\Delta} \{F(n)\} ,
\ee 
where ${\cal F}_{\Delta} \{F(n)\}$ is an operator notation for the Fourier
series transform of $F(n)$. 

%%%%%%%%%%%%%%%%%%%%%%%%%%%%%%%%%%%%%%%%%%%%%%%%%%%%%%%%%%%%%%%

The Fourier series transform ${\cal F}_{\Delta}$ of (\ref{C1})
gives (\ref{20}).
We will consider the limit $d \rightarrow 0$. 
Using (\ref{AR}), equation (\ref{20}) can be written as
\be \label{Eq-k}
\frac{\partial^2}{\partial t^2} \hat{u}(k,t) =
\frac{g_2 \, d^2}{M} \; \hat{\mathcal{T}}_{2, \Delta}(k) \; \hat{u}(k,t) +
\frac{g_{\alpha} d^{\alpha}}{M} \; \hat{\mathcal{T}}_{\alpha, \Delta}(k) \; \hat{u}(k,t)  
+ \frac{1}{M} \mathcal{F}_{\Delta} \{ F (n) \} , 
\ee
where we use (\ref{alpha-1}), the Proposition 1 for $K_2(n,m)$,  and the following notations
\be 
\hat{\mathcal{T}}_{\alpha , \Delta}(k) = 
- \frac{1}{2\Gamma(\alpha+1)} \, |k|^{\alpha} + 
d^2\, O(|k|^{\alpha +2})  ,
\ee
\be 
\hat{\mathcal{T}}_{2, \Delta}(k) = - k^2 +  d^2\, O(k^{4})  .
\ee

In the limit $d \rightarrow 0$, we get
\be 
\hat{\mathcal{T}}_{\alpha }(k) =
{\cal L} \, \hat{\mathcal{T}}_{\alpha , \Delta}(k) = - \frac{1}{2 \Gamma(\alpha +1)} \, |k|^{\alpha} ,
\ee
\be 
\hat{\mathcal{T}}_{2}(k) =
{\cal L}\hat{\mathcal{T}}_{2, \Delta}(k) = -  k^2 .
\ee
As a result, equation (\ref{Eq-k}) 
in the limit $d  \rightarrow 0$ gives
\be \label{Eq-k2}
\frac{\partial^2}{\partial t^2} \tilde{u}(k,t) =
\frac{g_2 \, d^2}{M} \; 
\hat{\mathcal{T}}_2(k) \; \tilde{u}(k,t) + 
\frac{g_{\alpha} d^{\alpha}}{M} \; \hat{\mathcal{T}}_{\alpha} (k) \; \tilde{u}(k,t)  + \frac{1}{M} \mathcal{F} \{ F (x) \}  ,
\ee
where $\tilde{u}(k,t)= {\cal L} \hat{u}(k,t)$ and $F(k)=
\mathcal{F} \{ F (x) \} = 
{\cal L} \, \mathcal{F}_{\Delta} \{ F (n) \} $. 
The inverse Fourier transform of (\ref{Eq-k2}) is
\be \label{Eq-x}
\frac{\partial^2}{\partial t^2} u(x,t) =
\frac{g_2 \, d^2}{M} \; \mathcal{T}_2(x) \; u(x,t) +
\frac{g_{\alpha} d^{\alpha}}{M} 
\; \mathcal{T}_{\alpha}(x) \; u(x,t) +
\frac{1}{\rho} f (x) ,
\ee
where $f(x)=F(x)/(A\, d)$ is the force density,
the operators $\mathcal{T}_2(x)$ and 
$\mathcal{T}_{\alpha}(x)$ are defined by
\be \label{Tx0}
\mathcal{T}_2(x) = 
\mathcal{F}^{-1} \{ \hat{\mathcal{T}}_2 (k) \} = 
\, \Delta , \quad
\mathcal{T}_{\alpha}(x) = 
\mathcal{F}^{-1} \{ \hat{\mathcal{T}}_{\alpha} (k) \} = 
 - \frac{1}{2 \Gamma(\alpha +1)} \, (- \Delta)^{\alpha/2} . 
\ee
Here, we use the connection between the Riesz's fractional 
Laplacian and its Fourier transform \cite{KST} in the form 
\be \label{FFL}
{\cal F}[ (-\Delta)^{\alpha/2} u(x)](k)= |k|^{\alpha} \, \hat u(k) .
\ee
Using the finite parameters $C_2$ and $C_{\alpha}$, 
which are defined by (\ref{G2G4}), 
the substitution of (\ref{Tx0}) into (\ref{Eq-x}) gives
continuum equations (\ref{CME}). 
This ends the proof.

%%%%%%%%%%%%%%%%%%%%%%%%%%%%%%%%%%%%%%%%%%%%%%%%%%%%%%%%%%%%%%
%%%%%%%%%%%%%%%%%%%%%%%%%%%%%%%%%%%%%%%%%%%%%%%%%%%%%%%%%%%%%%
%%%%%%%%%%%%%%%%%%%%%%%%%%%%%%%%%%%%%%%%%%%%%%%%%%%%%%%%%%%%%%

\end{document}